# A farewell to the MNCS and like size-independent indicators[1]

*Giovanni Abramo* (corresponding author)

Laboratory for Studies of Research and Technology Transfer
Institute for System Analysis and Computer Science (IASI-CNR)
National Research Council of Italy
ADDRESS: Istituto di Analisi dei Sistemi ed Informatica (IASI-CNR)
Via dei Taurini 19, 00185 Roma - ITALY
tel. and fax +39 06 72597362, giovanni.abramo@uniroma2.it

*Ciriaco Andrea D'Angelo*

Department of Engineering and Management - University of Rome "Tor Vergata" and
Institute for System Analysis and Computer Science (IASI-CNR)
ADDRESS: Università di Roma "Tor Vergata", Dip. di Ingegneria dell'Impresa
Via del Politecnico 1, 00133 Roma - ITALY
tel. and fax +39 06 72597362, dangelo@dii.uniroma2.it

**Abstract**

The arguments presented demonstrate that the Mean Normalized Citation Score (MNCS) and other size-independent indicators based on the ratio to publications are not indicators of research performance. The article provides examples of the distortions when rankings by MNCS are compared to those based on indicators of productivity. The authors propose recommendations for the scientometric community to switch to ranking by research efficiency, instead of MNCS and other size-independent indicators.

## 1. Introduction

While it may be debatable whether it was Albert Einstein or William Cameron that coined the saying, '*Not everything that can be counted counts, and not everything that counts can be counted*', no one doubts its pertinence and extraordinary popularity in the field of scientometrics. Our contention is that all size-independent indicators based on the ratio to publications, such as the Mean Normalized Citation Score (MNCS) and the rankings derived at any levels, in fact barely 'count'. Or what is worse, they may indeed count - but in the negative sense of leading to wrong decisions and policy.

The continuing drive for evidence-based decisions and policy-making in research systems has brought about a fervid search for what are justifiably hoped to be more precise, robust and reliable performance indicators and evaluation methods. Thus in recent years we have seen a proliferation of new indicators and variants or extensions of already famous ones, in particular the *h*-index (Hirsch, 2005). Bibliometricians are now running out of alphabet and subscript characters to name all the new indicators/variants. The drawbacks of the *h*-index have been discussed extensively in the literature, and there have been numerous attempts to overcome them through *h*-variants (Egghe, 2010; Norris and Oppenheim, 2010; Alonso et al., 2009). Instead, very little attention has been devoted to the validity of the CPP/FCSm or "old" crown indicator proposed by Leiden's CWTS (Van Raan, 2005; Moed et al., 1995), the MNCS or "new" crown indicator (Waltman et al., 2011), and all other size-independent indicators based on the ratio to publications. Apart from our own works (Abramo and D'Angelo, 2014; Abramo and D'Angelo, 2013), we find only one other study in the literature (Danell, 2013) expressing doubts about the validity of the MNCS and highly-cited articles out of total publications as indicators of performance. In this paper we focus on the MNCS and other like indicators, based on the ratio to publications. Its aim is to make it conclusively clear that these indicators are not at all indicators of research performance, and are not worthy of further use or attention.

## 2. The fallacy of the MNCS as indicator of performance

Several years ago in this same journal, we witnessed the tit for tat argument on the statistical normalization of the crown indicator between Opthof and Leydesdorff (2010; Leydesdorff and Opthof, 2011) and the bibliometricians from the CWTS group (Van Raan et al., 2010; Waltman et al., 2011). Other scholars joined the debate (Vinkler, 2012; Larivière and Gingras, 2011; Moed, 2010), and the argument ultimately led to a new definition of the crown indicator, the MNCS (Waltman et al. 2011)[2]. Several more scientometricians have assumed the validity of size-independent indicators based on the ratio to publications in their works: Ruiz-Castillo and Waltman (2015), Fairclough and Thelwall (2015), Aksnes et al. (2013), Glänzel et al. (2013), Bornmann et al. (2012), Zitt (2005), just to cite a few. However, common sense demonstrates that the MNCS and all other size-independent indicators based on the ratio to publications, whatever the technical details of their calculation, are invalid indicators of research performance. Thus, all of the many analyses based on these indicators should be revisited. The time and resources currently dedicated to improving them or employing them in research works could be more effectively devoted to the improvement of efficiency indicators. Ranking individuals or institutions by these indicators is at best of no value, and in fact presents serious dangers where they are used as the basis for decisions and policies.

---

[2] It was in fact Lundberg (2007) who first questioned the old crown indicator and proposed a new one, which the CWTS group then labeled as the MNCS.

The MNCS is claimed as an indicator of research performance, measuring the average number of citations of the publications of an individual or institution, normalized for subject category and publication year. Similarly, the share of individual or institutional publications belonging to the top 1% (10%, etc.) of 'highly cited articles' (HCAs), compared with other publications in the same field and year, is considered another indicator of research performance. For years we have seen the publication of international performance rankings by such size-independent indicators, based on the ratio to publications. Our contention is that research performance and relative rankings must if anything be drawn up by average field-normalized impact per euro spent on research or per researcher (preferably normalized by capital) and not per publication (Abramo and D'Angelo, 2014); or by HCAs per euro spent/researcher (Abramo and D'Angelo, 2015), and not HCAs out of publications.

Given two universities of the same size, resources and research fields, which one performs better: the one with 100 articles each earning 10 citations, or the one with 200 articles, of which 100 with 10 citations and the other 100 with five citations? A university with 10 HCAs out of 100 publications, or the one with 15 HCAs out of 200 publications? In both examples, by MNCS or proportion of HCAs, the second university performs worse than the first one (25% lower). But clearly, using common sense, the second is in both cases the better performer, as it shows higher returns on research investment (50% better). This is also the conclusion using our own proxy indicator of productivity: Fractional Scientific Strength (FSS), which embeds both quantity and impact[3]. Basic economic reasoning confirms that the better performer under parity of resources is the actor who produces more; or under parity of output, the better is the one who uses fewer resources. Indeed the MNCS, the proportion of HCAs, and all other size-independent indicators based on the ratio to publications are invalid indicators of performance, because they violate an axiom of production theory: as output increases under equal inputs, performance cannot be considered to diminish. Indeed an organization (or individual) will find itself in the paradoxical situation of a worsened MNCS ranking if it produces an additional article, whose normalized impact is even slightly below the previous MNCS value.

To give an idea of the distortions embedded in the MNCS-based rankings we provide a few examples, extracted from our regular analysis of performance in the Italian academic system.

First, for the period 2008-2012, we compare the performance ranking of over 36,000 professors in the sciences based on FSS to their ranking based on MNCS. Table 1 presents the results of the comparison for the top quartile of professors. On average, 42.7% of faculty that are 'top' by FSS would fail to reach this status when ranked by MNCS, with the share ranging from 32.1% in Civil Engineering to 48.5% in Earth Sciences.

Next, within these broad shifts, we show three extreme cases of differences in performance by MNCS and FSS at the individual, field, and discipline levels. Table 2 shows that Professor John Doe performs very poorly by MNCS, while colleague Jane

---

[3] For a number of years we have used this indicator to rank the performance of Italian universities and individuals (Abramo and D'Angelo, 2011; Abramo et al., 2010;). The definition and operationalization of the measure of FSS at individual and aggregate levels, may be found in Abramo and D'Angelo (2014).

Doe is at the top (100th percentile) of all scientists in her chosen field. But their relative positions are totally inverted in ranking by FSS. Also, in Table 3, we see that University A is the best in the nation by MNCS in the field of Diagnostic Imaging and Radiotherapy, but ranks at the bottom by FSS. Conversely, University B ranks last in Environmental and Applied Botanics by MNCS, but leaps to second by FSS. Finally Table 4 shows that University A ranks second out of 49 in the discipline of Industrial and Information Engineering when considered by MNCS, but only 47th by FSS. The opposite extremes occur for University B, whose percentile rank by MNCS in Physics is 33.3, but 97.6 by FSS.

*Table 1. Comparison of performance rankings by MNCS and FSS, for all Italian professors (based on 2008-2012 WoS publications)*

| UDA | Percentage of top 25% professors by FSS not included in the same set by MNCS |
|---|---|
| Mathematics and computer science | 33.4 |
| Physics | 44.6 |
| Chemistry | 40.8 |
| Earth sciences | 48.5 |
| Biology | 47.1 |
| Medicine | 46.6 |
| Agricultural and veterinary science | 41.6 |
| Civil engineering | 32.1 |
| Industrial and information engineering | 39.3 |
| Total | 42.7 |

*Table 2. Comparison of performance ranks by MNCS and FSS, for two professors of Chemistry*

| | Professor | John Doe | Jane Doe |
|---|---|---|---|
| | Field | Foundations of chemistry for technologies | Physical chemistry |
| | Number of publications | 70 | 1 |
| MNCS | Absolute value | 0.42 | 5.11 |
| | rank | 167 of 187 | 1 of 419 |
| | percentile | 10.8 | 100 |
| FSS | Absolute value | 3.39 | 0.01 |
| | rank | 6 of 187 | 402 of 419 |
| | percentile | 97.3 | 4.1 |

*Table 3. Comparison of performance ranks by MNCS and FSS, of two universities in two research fields*

| | University | A | B |
|---|---|---|---|
| | Field | Diagnostic imaging and radiotherapy | Environmental and applied botanics |
| | Number of publications | 15 | 65 |
| MNCS | Absolute value | 2.5 | 0.4 |
| | rank | 1 of 38 | 26 of 26 |
| | percentile | 100 | 0.0 |
| FSS | Absolute value | 0.1 | 1.7 |
| | rank | 38 of 38 | 2 of 26 |
| | percentile | 0.0 | 96.0 |

*Table 4. Comparison of performance ranks by MNCS and FSS, of two universities in two disciplines*

|  |  | University A | B |
|---|---|---|---|
|  | Discipline | Industrial and information engineering | Physics |
|  | Number of publications | 76 | 1008 |
| MNCS | Absolute value | 1.3 | 0.8 |
|  | rank | 2 of 49 | 29 of 43 |
|  | percentile | 97.9 | 33.3 |
| FSS | Absolute value | 0.4 | 1.9 |
|  | rank | 47 of 49 | 2 of 43 |
|  | percentile | 4.2 | 97.6 |

Given such discrepancies in rankings, the socio-economic consequences of MNCS-based decisions on selective funding, recruitment, promotion and turnover are easy to imagine. Yet the old crown indicator in the past, and now the MNCS, remain in use by almost the entire bibliometric community. An impressive number of studies rely on size-independent indicators based on the ratio to publications. Within our scientific community, the CWTS and SCImago have long put out international performance rankings based on these indicators. Elsewhere, the same indicators are used to draw up such highly popular university rankings as the THE[4] and US News[5] versions. The Science and Engineering Indicators report of the US National Science Foundation again use these same flawed tools.[6] Finally, such size-independent indicators play a prominent role in the two most popular commercial systems for bibliometric analysis: InCites by Thomson Reuters and SciVal by Scopus.

Why is it that bibliometricians would continue to resort to size-independent indicators based on the ratio to publications, producing rankings of no value to decision-makers? The answer is most likely that they do not know the numbers of scientists in the universities they consider. Whatever the reason, it is a degradation of our scientific service that these indicators in fact do not really measure performance, or live up to their vaunted claims. For instance: 'The CWTS Leiden Ranking 2015 offers key insights into the scientific performance of 750 major universities worldwide'[7].

On the other hand we can see that the public are so much in need of performance rankings that they have come to accept anything that comes with such a label, regardless of the content.

## 3. Recommendations

In this part of the paper we call for the scientometrics community and related stakeholders to expedite the shift to the new technological trajectory of efficiency indicators, and offer several strategic recommendations:
- First, we call all scientometricians interested in developing performance indicators and rankings to abandon the old and shift to the new research paradigm, dedicating attention exclusively to the improvement of indicators and methods for true measurement of research efficiency - such as productivity, including labor and total factor productivity, HCAs per research spending, and the like. Our willingness to accept this change depends on first accepting our responsibility, as scientists and citizens. We must resist the syndrome of 'publish or perish' (in the familiar topic of size-independent indicators), and instead work towards the advantages to be gained by new scientific progress.
- Second, we ask the chief editors and editorial boards of scientometric journals to play a leading role in the transition to the new research trajectory, through their editorial policies and selection of manuscripts. We make the same call to scientometricians, who act as reviewers of manuscripts for scientific journals.
- Our third call is to all those actors who produce performance rankings, whether scientometricians or not. Such organizations and individuals must have the honesty and courage to recognize that the current international bibliometric rankings are little more than mere arithmetic exercises, which fail to measure anything like what they were intended to: the performance of institutions. The existing rankings based on size-dependent indicators, such as an institution's total number of citations or total number of HCAs, do not permit discernment of how much a position is due to merit rather than to size. For those that are instead prepared using size-independent indicators based on the ratio to publications, we have revealed the inherent fallacies. For those rankings based on the h-index and its variants, each solving a single drawback but leaving others unsolved, a previous work described the embedding of distortions (Abramo et al., 2013).
- Our fourth call is to the governments and research institutions that intend to use any bibliometric performance rankings. If they expect ever more precise and reliable performance evaluations, useful for their decisions and policy making, then they must be prepared to give scientometricians the underlying data necessary for the job (i.e. name and affiliation of scientists, field of research, academic rank, resources allocated, etc.).

---

[4] https://www.timeshighereducation.com/world-university-rankings, last accessed on November 2, 2015.
[5] http://www.usnews.com/education/best-global-universities/rankings, last accessed on November 2, 2015.
[6] See "Trends in Citation of S&E Articles" in http://www.nsf.gov/statistics/seind14/content/chapter-5/chapter-5.pdf, last accessed on November 2, 2015.
[7] http://www.leidenranking.com, last accessed on November 2, 2015.

Finally, to develop increasingly distortion-free rankings, our recommendation is that not only do we adopt indicators of production efficiency but that we also classify the scientists by research field, to avoid distortions due to the different patterns of publication across fields (Abramo et al., 2008). In fact, if we all agree on normalizing citations by field to account for the different citation behavior across fields (for instance the 251 fields in the WoS classification), there is no reason to overlook the different intensity of publications across fields, and no justification for broad field aggregations when comparing performance. To give an idea of the distortions inherent in gross field classification of scientists, we extract an example from Italian faculties. In the Italian university system all professors are classified in one and only one field (370 in all), grouped into disciplines (14 in all). Table 5 presents the case of two professors, both belonging to the discipline of Clinical Medicine, more specifically one to the field of Blood diseases and the other to Vascular surgery. Without refined field classification, the direct comparison of the professors by any performance indicator would lead to the conclusion that the first one performs better (comparing the indicator values). The question is whether this result is due to the first professor's merit or to the higher intensity of publication and citation in the research field of Blood diseases. When a finer field classification is accounted for, and each professor's performance is first compared to that of their colleagues in the same field, the comparison of their positions in the respective field rankings (% ranks) reveals that the better performer is actually the second professor. What is clear is that for international performance comparisons based on distortion-free rankings, we need to arrive at a common international classification system of all scientific fields, therefore we are now calling on scientometricians to explore the paths to international coordination, for exactly this purpose.

*Table 5: Comparison of performance of two Italian professors, accounting for (Rank%) and not accounting for (Value) field classification. Data based on 2009-2013 WoS publications.*

| Name | John Doe | | Jane Doe | |
|---|---|---|---|---|
| Discipline | Clinical Medicine | | Clinical Medicine | |
| Field | Blood diseases | | Vascular surgery | |
| Indicator | Value | Rank% | Value | Rank% |
| O* | 6.6 | 67.4 | 3.6 | 90.5 |
| FO** | 1.4 | 68.4 | 1.2 | 95.2 |
| MNCS | 2.0 | 78.9 | 0.6 | 58.7 |
| FSS | 1.2 | 78.4 | 0.7 | 91.3 |
| h-index | 12 | 76.4 | 5 | 79.6 |

*\* O = yearly average output; \*\* FO = yearly average fractional output*

## 4. Conclusions

We are aware that many countries do not have exhaustive databases of the composition of their university faculties, and that the disambiguation of author names on national scale remains a difficult task. But these difficulties and challenges are no justification for us to continue scientometric research and development along a technological trajectory that had no foundations from the outset, and which cannot compare to true evaluation of research productivity.

It may take years before we are able to draw up global rankings based on efficiency indicators, but we know that these will ultimately be truly meaningful and useful to decision and policy makers. To attract ever more countries, we would be delighted to begin comparative analyses in collaboration with those scientometricians who already have access to exhaustive databases on the composition of their own national faculties. As the first comparisons are published, we expect a snowball effect would occur.

It may seem easier, more comfortable and safer to continue investing in what has worked before, holding onto the current research paradigm, defending vested interests and avoiding the inherent destruction of a paradigm shift, but this would be more risky and costly in the long run.

As Seneca said, '*Non quia difficilia sunt non audemus, sed quia non audemus difficilia sunt*', or '*It is not because things are difficult that we do not try; it is because we do not try that things are difficult*'.